\documentclass[a4paper,fleqn,usenatbib,useAMS]{mnras}

\usepackage[T1]{fontenc}
\usepackage{ae,aecompl}

\usepackage{graphicx}	
\usepackage{amsmath}	
\usepackage{amssymb}	
\usepackage{booktabs}   
\usepackage{mathrsfs}	
\usepackage{bm}         

\title[Estimating $P(k)$ covariance matrix]{Estimating the power spectrum covariance matrix with fewer mock samples}

\author[D. W. Pearson and L. Samushia]{
David W. Pearson$^{1}$\thanks{Email: dpearson@phys.ksu.edu} and
Lado Samushia$^{1,2,3}$
\\
$^{1}$Department of Physics, Kansas State University, 116 Cardwell Hall, Manhattan, KS 66506, USA\\
$^{2}$National Abastumani Astrophysical Observatory, Ilia State University, 2A Kazbegi Ave., GE-1060 Tbilisi, Georgia\\
$^{3}$Institute of Cosmology \& Gravitation, University of Portsmouth, PO1 3FX, UK 
}

\date{Accepted XXX. Received YYY; in original form ZZZ}

\pubyear{2015}

\defcitealias{Feldman1994}{FKP}

\begin{document}
\label{firstpage}
\pagerange{\pageref{firstpage}--\pageref{lastpage}}
\maketitle

\begin{abstract}
The covariance matrices of power-spectrum ($P(k)$) measurements from galaxy
surveys are difficult to compute theoretically. The current best practice is to
estimate covariance matrices by computing a sample covariance of a large number
of mock catalogues. The next generation of galaxy surveys will require thousands
of large volume mocks to determine the covariance matrices to desired accuracy.
The errors in the inverse covariance matrix are larger and scale with the number
of $P(k)$ bins, making the problem even more acute. We develop a method of
estimating covariance matrices using a theoretically justified, few-parameter
model, calibrated with mock catalogues. Using a set of 600 BOSS DR11 mock catalogues, 
we show that a seven parameter model is sufficient to fit the covariance matrix of 
BOSS DR11 $P(k)$ measurements. The covariance computed with this method is better 
than the sample covariance at any number of mocks and only $\sim$100 mocks are 
required for it to fully converge and the inverse covariance matrix converges at 
the same rate. This method should work equally well for the next generation of
galaxy surveys, although a demand for higher accuracy may require adding extra
parameters to the fitting function.
\end{abstract}


\section{Introduction}

The covariance matrix and inverse covariance matrix of the band averaged power
spectrum are crucial for parameter estimation from cosmological spectroscopic
surveys. Having an accurate estimate of the covariance matrix, and therefore the
inverse covariance matrix, is paramount in order to be able to assign reliable
uncertainties in estimated parameters  \citep{Percival2014}. Most studies
achieve this by using a large number of mock samples
\citep{Cole2005,Reid2010,Manera2013,Manera2015,Anderson2014,Gil-Marin2015} setup
to match the characteristics of the particular survey, and then running them
through the data pipeline to estimate the covariance matrix via
\begin{equation}
C_{ij} = \dfrac{1}{N-1}\sum_{s}\left(P_{i}^{s}-\mu_{i}\right)\left(P_{j}^{s}-\mu_{j}\right),
\label{eq:covar}
\end{equation}
where $N$ is the number of mocks,
\begin{equation}
\mu_{i} = \dfrac{1}{N}\sum_{s}P_{i}^{s},
\label{eq:mean}
\end{equation}
and \mbox{$P_i\equiv P(k_i)$}.

The elements of the sample covariance matrix converge as
$\mathcal{O}\left(N^{-1}\right)$ to their true values, while the inverse
covariance matrix elements converge as $\mathcal{O}\left(N_\mathrm{b}/N\right)$,
where $N_\mathrm{b}$ is the number of $P(k)$ bins \citep[see
e.g.,][]{AndersonBook}. This inaccuracy propagates into derived cosmological
parameters and inflates their errorbars by a factor of
\mbox{$\mathcal{O}\left(1+N_\mathrm{b}/N\right)$}
\citep{Hartlap2007,Taylor2013,Percival2014,DodelsonSchneider2013,TaylorJoachimi2014}.
\cite{Percival2014} found that in order for this extra variance to be
sub-dominant for Baryon Oscillation Spectroscopic Survey
\citep[BOSS;][]{Eisenstein2011,Dawson2013} Data Release (DR) 9 measurements
\mbox{$\sim 600$} mock catalogues were needed.  The next generation of galaxy surveys
will have more stringent requirements on the precision of the covariance matrix
and will require even larger sets of mock catalogues to compute the sample
variance. \citet{Taylor2013} estimated that up to $10^6$ mocks may be required
for the joint analysis of future galaxy clustering and weak lensing data.

There are a number of methods with which one can generate these mock catalogues
that tend to reduce the computational cost, such as the log-normal method
\citep{Coles1991}, pinpointing orbit-crossing collapsed heirarchical objects 
\citep[\textsc{pinocchio};][]{Monaco2002,Monaco2013}, comoving lagrangian
acceleration simulations \citep[\textsc{cola};][]{Tassev2013}, perturbation
theory catalogue generator of halo and galaxy distributions method
\citep[\textsc{patchy};][]{Kitaura2014,Kitaura2015}, pertabuation theory halos method
\citep[\textsc{pthalos};][]{Scoccimaro2002,Manera2013,Manera2015}, or quick particle mesh
method \citep[\textsc{qpm};][]{White2014}. \cite{Chuang2015} provide detailed
descriptions and a comparison of the effectiveness of mocks generated using
these techniques, concluding that the more efficient approximate solvers
can be used to reach a few per cent accuracy for clustering statistics on scales of
interest for large-scale structure analyses. 

However, with surveys such as the Dark Energy Survey
\citep[DES;][]{Frieman2013}, the upcoming Dark Energy Spectroscopic Instrument
survey \citep[DESI;][]{Schlegel2011,Levi2013}, Extended Baryon Oscillation
Spectroscopic Survey \citep[eBOSS;][]{Schlegel2009}, Large
Synoptic Survey Telescope surveys \citep[LSST;][]{LSST2009}, \textit{Euclid} satellite
mission surveys \citep{Laureijs2011}, and numerous others increasing
the volumes to be analysed, the mocks must also increase in volume.  This can
still lead to situations where the computational costs of generating the
necessary number of mocks becomes prohibitive, making it desirable to have
a method of estimating the true covariance matrix using fewer mock catalogues.

The use of mock catalogues can be completely bypassed by looking at the intrinsic scatter
of $P(k)$ measurements within the sample e.g. using the jack-knife or bootstrap methods.
The jack-knife method is limited by the fact that 
to build up better statistics one must divide the survey
data into smaller and smaller subvolumes, limiting the maximum scale for
which the covariance can be reliably measured \citep{Norberg2009,Beutler2011}.
The bootstrap method performs better but is still limited 
by the number of subvolumes that can be created from the data \citep[see][for a
detailed discussion of the two methods]{Norberg2009}.

In principle, the covariance of $P(k)$ measurements can be computed
theoretically. Nonlinear effects in structure growth and highly nontrivial
survey windows make this kind of computation difficult in practice. Despite
this, several recent works demonstrated that this approach can be used to derive
reasonably good approximations to the covariance matrix \citep[for recent work
see e.g.,][and references therein]{dePutter2012}. Similar efforts have been
applied to the correlation function (inverse Fourier transform of $P(k)$) covariance
matrices \citep[see e.g.,][]{Xu2012}.

Alternative approaches include using a shrinkage estimation
\citep{PopeSzapudi2008,delaTorre2013}, covariance tapering
\citep{PazSanchez2015} and using a small number of mocks while `resampling'
large-scale Fourier modes \citep{Schneider2011}.

In this paper we propose a new approach to estimating $P(k)$ covariance
matrices. We start with a brief overview of the theory behind $P(k)$ covariance
matrices in section~\ref{sec:theory}. Then we describe the mock catalogues along with
our procedure for estimating the power spectrum, true covariance and inverse 
covariance matrix, and their associated uncertainties, from those catalogues 
in section~\ref{sec:measurement}. 

In section~\ref{sec:calibrate},
we choose a theoretically justified functional form with a small number of free
parameters to describe the covariance matrix and use mock catalogues to
calibrate numerical values of parameters. Unlike previous approaches based on
theoretical modelling we put significantly less stress (and effort) into
computing the actual covariance matrix elements; `Back of the
envelope' theoretical considerations are only used as a rough guide in
justifying the functional form and the actual numbers come purely from the fit
to the mock sample covariance matrices. In section~\ref{sec:covarconv} we show that 
a simple seven parameter model is good enough to describe the covariance matrix
of the BOSS DR11 sample as computed from a sample of 600 mocks. In the range of
scales relevant for the baryon acoustic oscillation (BAO) peak and the redshift-space
distortion measurements (\mbox{$0 < k < 0.2h\ \mathrm{Mpc}$}) our fitting function works
exceptionally well.

We also show in section~\ref{sec:covarconv} that the procedure converges much better
than the sample covariance with the number of mocks used. At any number of mocks, 
the fitted covariance matrix is closer to the final result than the corresponding 
sample covariance matrix, and at \mbox{$N\sim100$} the fitted covariance matrix is already
statistically indistinguishable from the sample covariance matrix computed with 
\mbox{$N=600$}. The inverse covariance matrix converges at the same rate as the 
covariance matrix. Demand for higher accuracy may require introducing additional 
free parameters into the fitting function, but there is no reason why this method 
should not work equally well for the future galaxy surveys. These conclusions are summarized in 
section~\ref{sec:conclusion}, where we also discuss planned further work.

\section{Theoretical $P(k)$ covariance}
\label{sec:theory}

For a Gaussian field in a large uniform volume the covariance matrix of $P(k)$
estimated in bins of width $\delta k$ is
\begin{equation}
C_{ij} = \frac{(2\upi)^3}{V}\frac{\left(P_i + n^{-1}\right)^2}{2\upi k_{i}^{2}\delta k}\delta_{ij},
\end{equation}
where $V$ is the volume of the survey and $n$ is the number density of galaxies
\citep{Feldman1994,Tegmark1997}. When the number density of galaxies is not
constant across the survey this changes to
\begin{equation}
\label{eq:Cij}
C_{ij} = \frac{(2\upi)^3}{V_\mathrm{eff}(k_i)}\frac{P^2_i}{2\upi k_{i}^{2}\delta
k}\tilde{\delta}_{ij},
\end{equation}
with
\begin{equation}
V_\mathrm{eff}(k_i) \equiv
\displaystyle\int\displaylimits_V\!\mathrm{d}^3r\,\frac{\left[P^2_in(\bm{r})+1\right]^2}{P^2_in^2(\bm{r})}.
\end{equation}
If, in addition, the width of the k-bins is comparable to $1/V$, the finite
volume will result in the coupling of neighbouring $P(k)$ measurements and the
Kronecker delta function in Eq.~(\ref{eq:Cij}) will turn into
\begin{equation}
\delta_{ij} \rightarrow \tilde{\delta}_{ij} =
\displaystyle\int\displaylimits_V\!\mathrm{d}^3r\,\mathrm{e}^{-\mathrm{i}(\bm{k}_i-\bm{k}_j)\cdot\bm{r}}.
\end{equation}
The observed volume $V$ is usually highly nontrivial which makes the effective
volume difficult to compute.  Nonlinear effects in structure growth and galaxy
biasing further complicate matters, adding terms proportional to the bin averaged
trispectrum, \mbox{$\overline{T}(k_{i},k_{j})$}, to the off-diagonal elements of 
the covariance matrix, resulting in
\begin{equation}
\label{eq:covmat}
C_{ij} = \frac{(2\upi)^3}{V_\mathrm{eff}(k_i)}\left(\frac{P^{2}_{i}}{2\upi
k_{i}^{2}\delta k}\delta_{ij} + \overline{T}(k_i,k_j)\right)
\end{equation}
(see \citealt{Scoccimarro1999b} and \citealt{Bernardeau2002} for details). 

\section{Measuring $P(k)$ covariance from BOSS DR11 mocks}
\label{sec:measurement}

While it is possible, in principle, to compute the covariance of $P(k)$ measurements
from Eq.~\eqref{eq:covmat}, highly nontrivial survey windows and the introduction of
terms dependent on the trispectrum from nonlinear structure growth make this difficult
in practice. Therefore, in order to obtain an estimate of the true covariance matrix,
we use 600 BOSS DR11 \textsc{pthalos} mock catalogues \citep{Manera2013} to
compute the sample covariance of the spherically averaged $P(k)$. For simplicity we
only use the mocks for the North Galactic Cap (NGC). We use the same estimator,
weighting scheme, and shot noise subtraction method as the latest official DR11
analyses papers \citep[see e.g.,][]{Gil-Marin2015}. We estimate the spherically
averaged $P(k)$ in 23 bins of width \mbox{$\Delta k = 0.008h~\mathrm{Mpc}^{-1}$} in the
wavelength range \mbox{$0.0 \leq k \leq 0.184h~\mathrm{Mpc}^{-1}$}. We then compute the
sample covariance matrix using Eqs.~(\ref{eq:covar}) and (\ref{eq:mean}).

\begin{figure}
\includegraphics[width=\columnwidth]{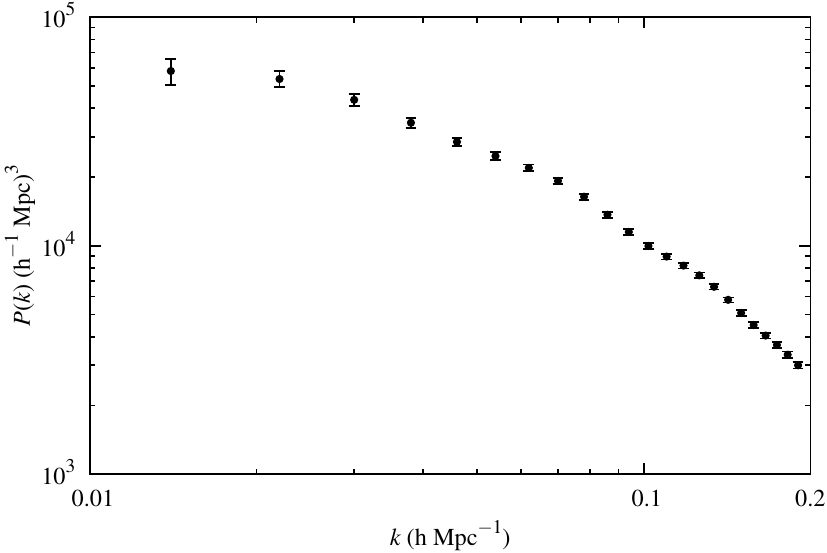}
\caption{The average of the power spectra from the 600 CMASS NGC \textsc{pthalos} mocks.
The error bars are the square root of the diagonal elements of the sample
covariance matrix calculated from all 600 mocks.}
\label{fig:avg_pk}
\end{figure}

\begin{figure}
\includegraphics[width=\columnwidth]{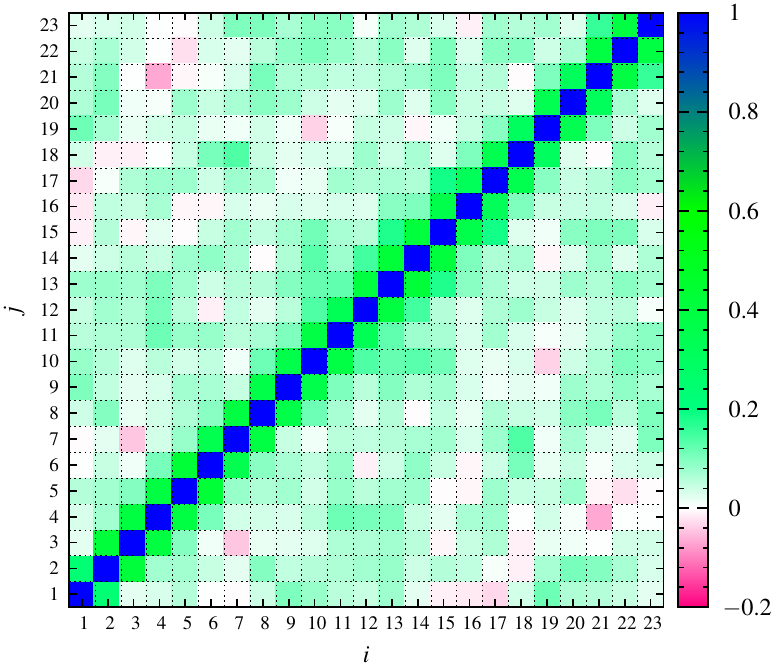}
\caption{The reduced covariance matrix calculated with all 600 NGC \textsc{pthalos} mocks. 
See the online article for a colour version of this plot.}
\label{fig:reducedcovar}
\end{figure}

Figure~\ref{fig:avg_pk} shows the average $P(k)$ with the errorbars computed by
taking a square root of diagonal elements of the covariance matrix
($\sqrt{C_{ii}}$). Figure~\ref{fig:reducedcovar} shows the elements of the
reduced covariance matrix defined as
\begin{equation}
r_{ij} \equiv \dfrac{C_{ij}}{\sqrt{C_{ii}C_{jj}}}.
\label{eq:correlationmatrix}
\end{equation}

Assuming that the distribution of individual $P(k)$ measurements is close to
Gaussian, the measured $C_{ij}$ follow the Wishart distribution
\citep{AndersonBook}. In the limit of large $N$ the Wishart distribution tends to a
Gaussian distribution with the covariance matrix 
\begin{equation}
\langle C_{ij}C_{kl} \rangle=\frac{\sigma_i\sigma_j\sigma_k\sigma_l}{N}(r_{ik}r_{jl} +
r_{il}r_{jk}),
\end{equation}
where $\sigma_i$ and $r_{ij}$ are the unknown true variance and cross-correlation
coefficients of power spectrum band estimates.
The variance in both diagonal,
\begin{equation}
\label{eq:vardiag}
\langle C_{ii}^2 \rangle = \frac{2\sigma_i^4}{N},
\end{equation}
and off-diagonal,
\begin{equation}
\label{eq:varoffdiag}
\langle C_{ij}^2 \rangle = \frac{\sigma_i^2\sigma_j^2}{N}(1+r_{ij}^2),
\end{equation}
elements of the covariance matrix, estimated using Eq.~(\ref{eq:covar}), scale
with the number of mocks, $N$.
The cross-correlation between estimates of different $C_{ij}$ elements is of
order of $r_{ij}^2$ and can be safely ignored as the measured $r_{ij}$ are of
order of 0.01 or less.

Of course, the inverse covariance matrix is the quantity of interest for parameter 
estimation. While we are using an unbiased estimator to determine our sample covariance, the 
inverse of the sample covariance will not, in general, be unbiased
\citep[see][for details]{Hartlap2007}. In order to obtain an unbiased estimate of the true inverse 
covariance matrix it is necessary to apply a correction of the form
\begin{equation}
\label{eq:inverseestimate}
\Psi = \dfrac{N-N_{\mathrm{b}}-2}{N-1}\mathsf{C}^{-1}
\end{equation}
where $N_{\mathrm{b}}$ is the number of $P(k)$ bins \citep{Hartlap2007}. 
The variance in the elements of this unbiased inverse covariance 
matrix can be found as (see e.g., \citealt{Taylor2013}, \citealt{Percival2014} 
and references therein)
\begin{equation}
\label{eq:varinverse}
\langle \Delta\Psi_{ij}\Delta\Psi_{kl} \rangle = A\Psi_{ij}\Psi_{kl}+B(\Psi_{ik}\Psi_{jl} + \Psi_{il}\Psi_{jk})
\end{equation}
where
\begin{equation}
\label{eq:Acoeff}
\begin{array}{@{}l}
A = \dfrac{2}{(N-N_{\mathrm{b}}-1)(N-N_{\mathrm{b}}-4)}, \\ [2ex]
B = \dfrac{(N-N_{\mathrm{b}}-2)}{(N-N_{\mathrm{b}}-1)(N-N_{\mathrm{b}}-4)}.
\end{array}
\end{equation}
This leads to a variance in the diagonal elements of
\begin{equation}
\label{eq:varinvdiag}
\langle \Delta\Psi_{ii}^{2} \rangle = (A+2B)\Psi_{ii}^{2},
\end{equation}
and the off-diagonal elements of 
\begin{equation}
\label{eq:varinvnondiag}
\langle \Delta\Psi_{ij}^{2} \rangle = (A+B)\Psi_{ij}^{2}+B\Psi_{ii}\Psi_{jj}.
\end{equation}

To summarize, we estimate a 23 by 23 covariance matrix of $P(k)$ measurements
using 600 mocks. The errors on the 276 independent elements of the sample covariance
matrix are given by Eqs.~(\ref{eq:vardiag}) and (\ref{eq:varoffdiag}) with
negligible cross-correlation. In section~\ref{sec:calibrate} we will use the sample
$C_{ij}$ elements and their errorbars estimated in this way to calibrate the
parameters of the theoretical covariance matrix. Additionally, we estimate the inverse
sample covariance matrix using Eq.~\eqref{eq:inverseestimate}, and obtain errors on
the independent elements with Eqs.~\eqref{eq:varinvdiag} and \eqref{eq:varinvnondiag}. 
We will use these in  section~\ref{sec:covarconv}, to compare how the sample and theoretical 
inverse covariance matrices converge.

\section{Calibrating parameters of theoretical covariance matrix}
\label{sec:calibrate}

Nontrivial survey volume and difficulties inherent in the theory of cosmological
structure growth in the nonlinear regime make direct computation of the covariance
matrix in Eq.~(\ref{eq:covmat}) a highly nontrivial task. Much effort has been put 
into understanding the trispectrum of the galaxy field in translinear and nonlinear 
limits \citep[see e.g.][]{Scoccimarro1999a,SefusattiScoccimarro2005}. Here we will attempt
to construct a relatively simple, few-parameter function to approximate
the true covariance matrix. This fitting function will, of course, be a very crude
approximation to the true structure of the trispectrum. However, for the BOSS DR11 mocks used
here, it seems to achieve desirable accuracy over a wavelength range 
relevant to BAO analysis. 

We start by defining two functions, $f(k)$ -- to describe the behaviour of the
diagonal elements, and $g(k)$ -- to describe the behaviour of the off-diagonal
elements in the correlation matrix. These functions are defined through
\mbox{$C_{ii}(k_{i})~=~P_{i}^{2}f^{2}(k_{i})$} and \mbox{$r_{ij} = g(k_{i} - k_{j})$}. 
The first equation is the definition of $f(k_i)$, while the second equation implies that the
reduced covariance matrix depends only on the difference between the centres of
bins, an assumption that, in general, does not have to hold. 

$f(k)$ is a fractional error in the $P(k)$ measurement and since the sample is
weighted in such a way as to optimize the $P(k)$ measurement at BAO scales we
expect it to be a smooth function with a minimum around those scales. We adopt
a three-parameter function 
\begin{equation}
\label{eq:ffin}
f(k) = (ak)^b \mathrm{e}^{\nu k},
\end{equation}
which we justify later in this section.

The off-diagonal elements are generated by the window function effects
($V_\mathrm{eff}$) and the trispectrum ($\overline{T}$). For a simple case of
a uniform sample within a cubic volume and no additional selection effects the
cross-correlation is
\begin{equation}
\tilde{\delta}_{ij} \rightarrow g(\Delta k) \propto \mathrm{sinc} (\omega\Delta k),
\end{equation}
where \mbox{$\omega = L/2\upi$}, $L$ is the size of the cube, and \mbox{$\mathrm{sinc}(x) = \sin(x)/x$}.
Non-linear gravitational effects will induce some coupling of $k$-modes near the
diagonal \citep{Meiksin1999,Scoccimarro1999b,Sefusatti2006}  which we model 
by a Lorentzian function
\begin{equation}
g(\Delta k)  \propto \frac{\gamma^2}{(\Delta k)^2 + \gamma^2}.
\end{equation}
More subtle effects such as the `beat-coupling'
\citep{Hamilton2006,Rimes2006,Sefusatti2006} and `local average'
\citep{Sirko2005,Takahashi2009,dePutter2012} will result in a cross-correlation
even for large $\Delta k$. To account for those, we add a constant term to
$g(\Delta k)$.  By combining the above effects we get 
\begin{equation}
\label{eq:lorentziansinc}
g(\Delta k) = \frac{\alpha\left[\gamma^{2}/(\Delta k^{2}+\gamma^2)\right]+
(1-\alpha)\mathrm{sinc}(\omega\Delta k) + \beta}{1+\beta}.
\end{equation}
The terms are combined in such a way as to enforce $g(0) = 1$.

Our final ansatz for the covariance matrix is 
\begin{equation}
\label{eq:ansatz}
C_{ij}=P_iP_j f(k_i)f(k_j)g(k_i - k_j)
\end{equation}
with functions $f(k)$ and \mbox{$g(k_i - k_j)$} given by Eqs.~(\ref{eq:ffin}) and
(\ref{eq:lorentziansinc}) respectively. We will find the best-fitting numerical values for
free parameters $a$, $b$, $\nu$, $\alpha$, $\gamma$, $\omega$, and $\beta$, by fitting
this ansatz to the sample covariance matrix measured from 600 mocks.

\begin{figure}
\includegraphics[width=\columnwidth]{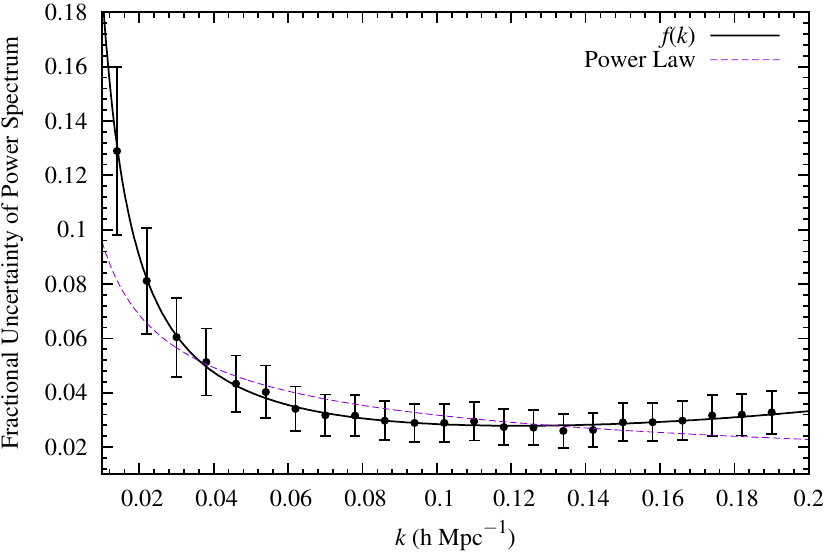}
\caption{The scaled diagonal elements of the covariance matrix determined from
all 600 mocks as a function of $k$. With the chosen scaling this is the
fractional uncertainty of the power spectrum. Our fitting function (\mbox{$f(k) =
(ak)^b\mathrm{e}^{\nu k}$}) follows the trend of the data remarkably well, while
fitting functions with fewer free parameters (power law -- \mbox{$f(k) = (ak)^b$})
cannot model the shape accurately.}
\label{fig:diagfit}
\end{figure}

Figure~\ref{fig:diagfit} shows $\sqrt{C_{ii}}/P_{i}$ (the fractional
uncertainty in $P(k)$ computed from the mocks) where the errorbars are computed using
Eq.~(\ref{eq:vardiag}). Our fitting function provides an excellent fit to its
shape. We find that the shape is difficult to recreate using functions with
fewer free parameters, such as a simple power law.

\begin{figure}
\includegraphics[width=\columnwidth]{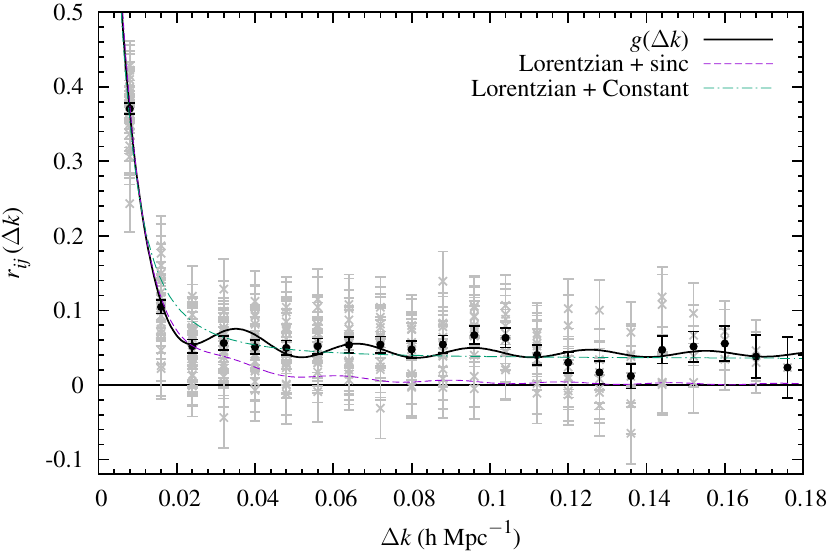}
\caption{The independent off-diagonal elements of the correlation matrix plotted as a
function of $\Delta k$. Grey data points represent individual $r_{ij}$ elements
while the black data points are the weighted mean and variance for the elements
with the same value of \mbox{$k_i - k_j$}. Our proposed function, $g(\Delta k)$, seems
to fit quite well for the full range of $\Delta k$ (\mbox{$\chi^{2} = 219.6$}, \mbox{$N_{\mathrm{DoF}} = 249$}). 
The best-fits for the Lorentzian plus constant term and Lorentzian plus sinc term 
result in noticeably worse fits (\mbox{$\chi^{2} = 252.9$}, \mbox{$N_{\mathrm{DoF}} = 251$} and 
\mbox{$\chi^{2} = 388.7$}, \mbox{$N_{\mathrm{DoF}} = 250$}, respectively). Note that the 
terms of each fitted function are combined in such a way as to enforce that they are equal to one 
when \mbox{$\Delta k = 0$}.}
\label{fig:offdiagfit}
\end{figure}

Figure~\ref{fig:offdiagfit} shows a similar plot for the off-diagonal elements of
the reduced covariance matrix, where we plotted the measurements in terms of
$\Delta k$. Our fitting function seems to provide a good phenomenological
description of this function. We find that reducing the number of parameters by
eliminating either the sinc term (by setting $\alpha = 1$) or the $\beta$ term
(by setting $\beta = 0$) significantly worsens the fit.

After performing the full fit to all independent covariance matrix elements we
get \mbox{$a = (451 \pm 35)h^{-1}\mathrm{Mpc}$}, \mbox{$b = -1.19 \pm 0.02$}, 
\mbox{$\nu = (9.62 \pm 0.32)h^{-1}\mathrm{Mpc}$}, \mbox{$\alpha = 0.867 \pm 0.024$},
\mbox{$\gamma = (5.17 \pm 0.16)\times 10^{-3} h~\mathrm{Mpc}^{-1}$}, \mbox{$\omega = 
(211.35 \pm 6.14)h^{-1}\mathrm{Mpc}$}, and \mbox{$\beta = 0.0423 \pm 0.0033$}, with \mbox{$\chi^2 = 250.6$}
for 269 degrees of freedom. The best-fitting value for $\omega$ is close to the
theoretically expected value of the average depth of the survey divided by $2\upi$.

\section{Convergence of the covariance matrix}
\label{sec:covarconv}
\begin{figure}
\includegraphics[width=\columnwidth,natheight=2650,natwidth=1983]{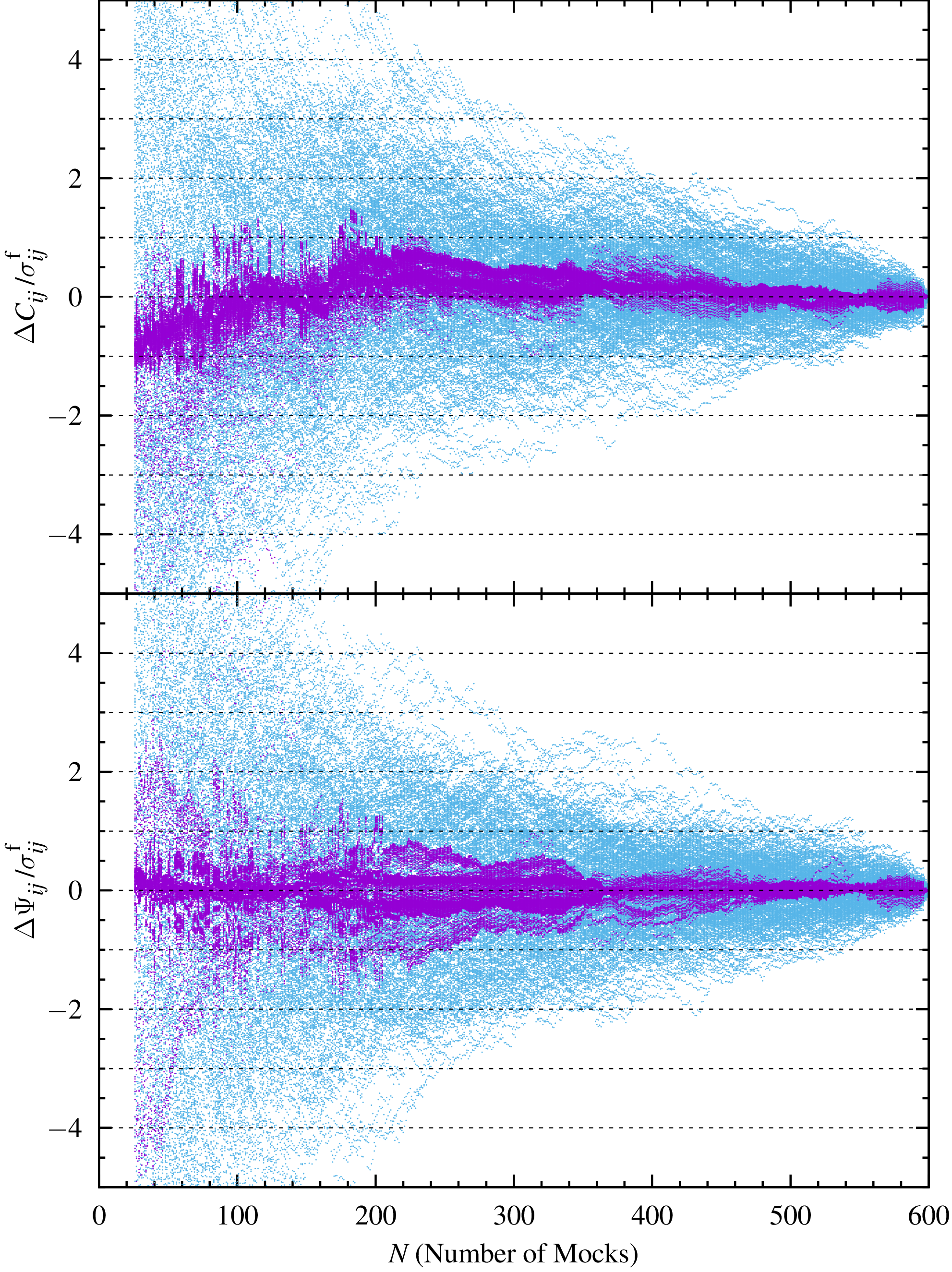}
\caption{Convergence of the covariance (top) and inverse covariance (bottom)
matrices to their final values (computed using 600 mocks) with the number of
mocks. The light blue points show the matrices determined from the mock catalogs
in the standard way, while the purple dots show the matrices as determined using
our fitting function. It is clear that the matrices from the best-fitting
functions converge much faster than the ones determined from sample variance.
See the online article for the colour version of this plot.}
\label{fig:convergence}
\end{figure}

The main advantage of the fitting function approach is that the covariance
matrix elements converge to their true values much faster than the sample
variance. Figure~\ref{fig:convergence} shows the offset of individual covariance
(and inverse covariance) matrix elements estimated using sample variance (light blue
points) and our method (purple points) as the number of mocks increases. The
offset is normalized to the standard deviation from the final (\mbox{$N=600$})
covariance (and inverse covariance) matrix given by Eqs.~\eqref{eq:vardiag} and 
\eqref{eq:varoffdiag} (or Eqs.~\eqref{eq:varinvdiag} and \eqref{eq:varinvnondiag} 
for the inverse). The fitting function method converges to the final result
much faster both for the covariance and inverse covariance matrices.

\begin{figure}
\includegraphics[width=\columnwidth]{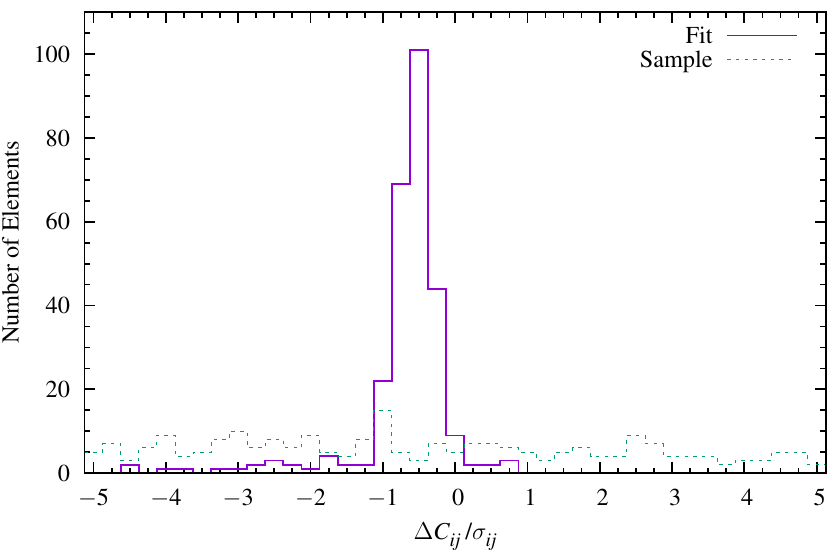}
\caption{Histograms of the differences between intermediate (generated using
first 50 mocks) covariance matrices and the final (using all 600 mocks)
covariance matrix normalized by the uncertainty in the final sample covariance
matrix. While slightly biased, the fitting functions already provide a significantly
better approximation of the final covariance matrix.}
\label{fig:covarsighist}
\end{figure}

Figure~\ref{fig:covarsighist} shows a histogram of the distribution of estimated
$C_{ij}$ elements around their true value for \mbox{$N=50$}. This is a horizontal slice
of the top panel of Figure~\ref{fig:convergence}. Already, very few elements
estimated with the fitting function method are outside 3$\sigma$ of the true
covariance, while for the sample variance method the distribution is basically
flat.

At low $N$, there is a small bias in $C_{ij}$ elements determined from the fitting function
method but it's significantly smaller than the variance in $C_{ij}$-s determined
from the sample variance and therefore the fitting function method is still
superior.

\begin{figure}
\includegraphics[width=\columnwidth]{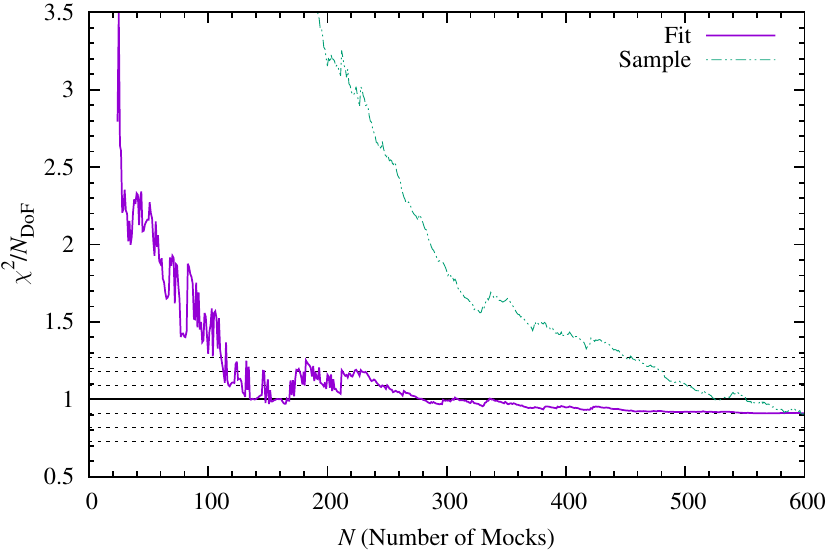}
\caption{Reduced $\chi^{2}$ of the covariance matrix compared to the final
covariance. The solid line is for the results obtained using the
fitting function and the dot-dashed line is for the sample covariance matrix.
Horizontal dashed lines show the expected 1, 2, and 3$\sigma$ deviations from the
\mbox{$\chi^2/N_{\mathrm{DoF}} = 1$} line.}
\label{fig:covarchisquare}
\end{figure}

Figure~\ref{fig:covarchisquare} shows the convergence to the final covariance in
terms of $\chi^2$ per degree of freedom (DoF) for the two methods. Since the
final covariance matrices computed with fitting function and sample variance
methods are close to each other in $\chi^2$, either one of them can be treated as
a good approximation for the true covariance matrix. To put both methods on
equal footing, when computing the $\chi^2$ for the fitting function method we
compare it to the final (\mbox{$N=600$}) sample variance method and
vice versa. Figure~\ref{fig:covarchisquare} clearly demonstrates that the
fitting function method generated $C_{ij}$-s become consistent with the true
covariance matrix much faster (already at \mbox{$N\sim100$}) compared to those generated
from the sample variance.

\section{Conclusions}
\label{sec:conclusion}

We propose a new method of estimating the elements of the covariance matrix for
band-averaged $P(k)$ measurements from galaxy surveys. The essence of the method
is to find a fitting function for the covariance matrix and calibrate its
parameters using a sample covariance computed from a set of mock catalogues. We
show that for the $P(k)$ measurements from the BOSS DR11 data in the range of 
\mbox{$0 \leq k \leq 0.2h~\mathrm{Mpc}^{-1}$} a very simple, seven-parameter function, given by
Eq.~\eqref{eq:ansatz}, is sufficient to describe the covariance matrix.  
The fitting function covariance matrix is statistically indistinguishable from the sample
covariance matrix computed from 600 mocks catalogues. 

The greatest advantage of this method, compared to the standard method of using
the sample covariance matrix, is that it requires significantly fewer mock catalogues
for calibration to converge to the true covariance matrix (see
figure~\ref{fig:convergence}). For the BOSS DR11 data, the fitting function
generated covariance matrices calibrated with as few as $\sim$100 mocks were statistically
indistinguishable from the sample covariance matrix generated with 600 mocks. To
get a similar convergence with the sample covariance matrix we had to use more
than 400 mocks (see figure~\ref{fig:covarchisquare}). This advantage of the new
method is especially relevant in situations where only a few mock catalogues are
available. With only 50 mock catalogues the distribution of the sample covariance
around the true value is basically flat, while the fitting function method already
provides a decent approximation (see figure~\ref{fig:covarsighist}).

The specific functional form that we tested in this work may turn out to be an
approximation that is too crude for future surveys, which will demand much higher
accuracy on the determination of $C_{ij}$. The functional form may have to be
modified if one wishes to include $P(k)$ measurements on much smaller scales as
well. However, once an appropriate (for the sought precision) functional form is
found, there is no reason why this method should not work with future data.

In future work we plan to study how this new method works at higher precision.
The uncertainties in the elements of the sample covariance (and inverse sample
covariance) matrix scale inversely with the number of mocks
(see seciont~\ref{sec:measurement}). A larger suite of mock catalogues would enable
us to better estimate the true covariance (and inverse covariance) matrix with
much smaller error bars on its elements. This would allow us to determine what
modifications of our fitting function are required at higher precision
to model the true covariance matrix.  This would have the additional benefit of
testing the fitting function method against a set of mocks which is completely
independent from the set used in this work, ruling out the possibility that our
successes were the result of some peculiarity which may be present in the data.  

This work was concerned only with the band-averaged $P(k)$. Higher order
Legendre polynomials of $P(k)$ with respect to the line-of-sight provide 
valuable information about the nature of gravity and the expansion rate. The
measurement of various Legendre moments of $P(k)$ will be cross-correlated. We
will address the question of modelling this larger covariance (and
inverse covariance) matrix in future work as well.

\section*{Acknowledgements}

LS is grateful for the support from SNSF grant SCOPES IZ73Z0 152581, GNSF
grant FR/339/6-350/14. This work was supported in part by DOE grant DEFG 03-99EP41093. 
NASA's Astrophysics Data System Bibliographic Service and the arXiv e-print service were 
used for this work. Additionally, we wish to acknowledge gnuplot\footnote{gnuplot 5.0 
was used in this work and can be downloaded at \href{http://www.gnuplot.info}{www.gnuplot.info}.}, 
a free open-source plotting utility which was used to create all of our figures and 
perform the function fitting central to this work. Lastly, we wish to acknowledge the many
useful conversations with Bharat Ratra while performing this work.


\bibliographystyle{mnras}
\bibliography{CovarMatrixEst}

\bsp	
\label{lastpage}
\end{document}